\documentclass[runningheads]{llncs}
\usepackage{algorithm2e}

\usepackage{xcolor}

\usepackage{amsmath, amssymb}
\usepackage{amsthm}

\usepackage[T1]{fontenc}
\usepackage{graphicx}
\begin{document}
\title{A Note on Efficient Privacy-Preserving Similarity Search for Encrypted Vectors}

\author{Dongfang Zhao} 
\institute{
University of Washington, USA \\
\email{dzhao@cs.washington.edu}
}

\maketitle              % typeset the header of the contribution

\begin{abstract}
Traditional approaches to vector similarity search over encrypted data rely on fully homomorphic encryption (FHE) to enable computation without decryption. However, the substantial computational overhead of FHE makes it impractical for large-scale real-time applications.
This work explores a more efficient alternative: using additively homomorphic encryption (AHE) for privacy-preserving similarity search. We consider scenarios where either the query vector or the database vectors remain encrypted, a setting that frequently arises in applications such as confidential recommender systems and secure federated learning. While AHE only supports addition and scalar multiplication, we show that it is sufficient to compute inner product similarity—one of the most widely used similarity measures in vector retrieval. Compared to FHE-based solutions, our approach significantly reduces computational overhead by avoiding ciphertext-ciphertext multiplications and bootstrapping, while still preserving correctness and privacy. We present an efficient algorithm for encrypted similarity search under AHE and analyze its error growth and security implications. Our method provides a scalable and practical solution for privacy-preserving vector search in real-world machine learning applications.
\end{abstract}

\section{Introduction}

Vector similarity search is a fundamental operation in modern machine learning and information retrieval applications, particularly in vector databases~\cite{solmaz_icict24} used for recommendation systems, retrieval-augmented generation (RAG), and large-scale search~\cite{milvus2021}. A key challenge in these applications is preserving data privacy while maintaining efficient query processing~\cite{otawose_sigmod23}. Fully Homomorphic Encryption (FHE) has been widely explored for encrypted similarity search but remains computationally expensive, making it impractical for real-time large-scale applications~\cite{cgentry_stoc09,bgv,bfv,ckks}. Open-source implementations such as Microsoft SEAL~\cite{sealcrypto} and IBM HElib~\cite{helib} provide FHE-based secure computation, but their high computational overhead significantly limits practical scalability~\cite{mkhan_icws24}.

In many privacy-sensitive applications~\cite{prism_sigmod21,otawose_sigmod23}, only one component—either the stored database vectors or the query vector—requires encryption, but not both. This occurs naturally in settings such as secure federated learning, confidential recommendation systems, and privacy-preserving search. The ability to perform similarity computations in such settings without decrypting the data is critical for balancing privacy and efficiency. Existing secure computation frameworks, such as SecureML~\cite{mohassel2017secureml}, provide strong privacy guarantees for machine learning tasks but often introduce significant computational overhead, making them less suitable for large-scale vector retrieval.

We investigate the feasibility of performing vector similarity search using only additively homomorphic encryption (AHE), which supports addition and scalar multiplication in the encrypted domain. Unlike FHE, which enables arbitrary computations at a high cost, AHE provides a significantly more efficient alternative by allowing direct computation of inner products without expensive ciphertext-ciphertext multiplications or bootstrapping. Despite its limited functionality, we demonstrate that inner product similarity—one of the most widely used metrics in vector retrieval—can still be efficiently computed under AHE, making it a viable choice for secure and scalable encrypted search.

This work shows that meaningful similarity search can be performed efficiently even when only addition is supported homomorphically. By leveraging AHE, we achieve substantial computational savings compared to traditional fully homomorphic methods while preserving correctness and privacy. These results suggest that lightweight homomorphic encryption schemes can play a crucial role in privacy-preserving vector search, providing a practical and scalable alternative to existing solutions.

\section{Preliminaries and Related Work}

\subsection{Vector Databases}
Vector databases have gained significant attention with the rise of applications such as natural language processing, recommendation systems, and information retrieval, where large-scale vector similarity searches are fundamental. Various systems have been designed to support these needs. For instance, Pinecone~\cite{pinecone} and Vearch~\cite{vearch} provide efficient vector search capabilities but often lack flexibility in handling non-vector queries or supporting mixed workloads. These systems primarily rely on approximate nearest neighbor (ANN) search techniques, with graph-based methods such as Hierarchical Navigable Small Worlds (HNSW)~\cite{hnsw} being widely adopted for scaling large datasets~\cite{johnson2019}.

Among purpose-built vector database management systems, Milvus~\cite{milvus2021} distinguishes itself by supporting multiple index types and hybrid queries, allowing structured attributes to be combined with vector-based similarity searches. This flexibility enables Milvus to integrate traditional query optimization techniques with emerging data modalities, including unstructured text and images. Other systems, such as Weaviate~\cite{weaviate2021} and SingleStore-V~\cite{singlestorev}, further enhance vector search capabilities by incorporating features like predicate-based vector search and cost-based optimization, making them suitable for federated and distributed environments.

Despite these advancements, many existing solutions lack efficient mechanisms for preserving privacy in federated learning and distributed systems. This limitation is particularly critical for real-time applications, where cryptographic protocols often introduce significant computational overhead~\cite{otawose_sigmod23}. While prior work in privacy-preserving machine learning, such as SecureML~\cite{mohassel2017secureml}, establishes foundational approaches, our protocol aims to enhance vector data management by integrating secure and efficient cryptographic mechanisms. These improvements ensure both high-performance execution and strong privacy guarantees in federated environments.

\subsection{Similarity Metrics}

Vector similarity measures play a crucial role in applications such as retrieval-augmented generation, recommendation systems, and large-scale search. Various similarity metrics have been developed, each tailored to different types of data distributions and retrieval objectives.

\paragraph{Cosine Similarity} One of the most widely used similarity metrics, cosine similarity measures the cosine of the angle between two vectors. Given two vectors $\mathbf{x}, \mathbf{y} \in \mathbb{R}^d$, it is computed as:

\[
    \text{CosSim}(\mathbf{x}, \mathbf{y}) = \frac{\mathbf{x} \cdot \mathbf{y}}{\|\mathbf{x}\| \|\mathbf{y}\|}.
\]

Cosine similarity is invariant to the magnitude of vectors, making it particularly effective for text embeddings and high-dimensional spaces where relative orientation is more informative than magnitude.

\paragraph{Inner Product} Inner product similarity directly measures the dot product between two vectors:

\[
    \text{InnerProd}(\mathbf{x}, \mathbf{y}) = \mathbf{x} \cdot \mathbf{y} = \sum_{i=1}^{d} x_i y_i.
\]

This measure is commonly used in machine learning models and recommendation systems, where higher values indicate greater relevance or similarity. It is particularly useful in neural network-based retrieval tasks.

\paragraph{Euclidean Distance} Euclidean distance quantifies the straight-line distance between two vectors:

\[
    \text{EucDist}(\mathbf{x}, \mathbf{y}) = \|\mathbf{x} - \mathbf{y}\| = \sqrt{\sum_{i=1}^{d} (x_i - y_i)^2}.
\]

Unlike cosine similarity, Euclidean distance is sensitive to vector magnitude and is often preferred in metric spaces where absolute differences matter, such as in image retrieval or clustering applications.

\paragraph{Learned Similarity Metrics} Recent advances in similarity search have explored learned similarity functions that adapt retrieval behavior beyond traditional metrics. The work by Ding et al.~\cite{bding_www25} introduces a framework that learns task-specific similarity metrics to improve retrieval performance. Unlike predefined measures such as cosine or Euclidean distance, their approach optimizes a similarity function through machine learning, enabling more adaptive and context-aware retrieval in large-scale vector search.

\subsection{Algebraic Homomorphim}

The notion \textit{homomorphism} refers to a class of functions that preserve the algebraic structures of the input and output spaces.
More specifically, an algebraic group\footnote{We use \textit{algebraic group} to refer to a group structure in group theory, \textit{not} the solutions to a system of polynomial equations in algebraic geometry.} can be relabelled and transformed, through a homomorphic function, into another algebraic group without changing the relationship among the elements. 
An algebraic \textit{group} is defined a nonempty set of elements along with a binary operator satisfying the closure, associativity, identity, and inverse properties.

Formally, a group $G$ over a set $S$ is represented as a tuple $(G, \oplus)$, where $\oplus$ is a binary operator. This operator adheres to four fundamental axioms or properties, expressed through first-order logical formulas:
(i) For all $g, h \in S$, $g \oplus h \in S$;
(ii) There exists a unique element $u \in S$ such that for all $g \in S$, $(g \oplus u = g)$ and $(u \oplus g = g)$;
(iii) For every $g \in S$, there exists an element $h \in S$ such that $(g \oplus h = u)$ and $(h \oplus g = u)$, where $h$ is often denoted as $-g$; and
(iv) For all $g, h, j \in S$, $(g \oplus h) \oplus j = g \oplus (h \oplus j)$.
Given another group $(H, \otimes)$ and a function $\varphi: G \rightarrow H$ that for all $g_1, g_2 \in G$ satisfies $\varphi(g_1) \otimes \varphi(g_2) = \varphi(g_1 \oplus g_2)$, we call function $\varphi$ a homomorphism from $G$ to $H$. This implies that $\varphi$ preserves the group operation between the elements of $G$ when mapped to the corresponding elements in $H$.

\section{Method}

\subsection{Definitions and Notations}

Given a plaintext vector $\mathbf{x} = (x_1, x_2, \dots, x_d)$ and an encrypted vector $\textbf{y}' = Enc(\mathbf{y}) = (Enc(y_1), Enc(y_2), \dots, Enc(y_d))$, where each $y_i$ is encrypted under an additively homomorphic cryptosystem, our goal is to compute an encrypted form of their similarity (i.e., inner product) $s'$:
\[
s' \equiv Enc\left( \sum_{i=1}^{d} x_i y_i \right)
\]
up to encryption randomness without revealing $\mathbf{y}$, such that $Dec(s') = s = \sum_{i=1}^{d} x_i y_i$.

Let's define an additively homomorphic encryption scheme formally:

\begin{definition}[Additively Homomorphic Encryption]
An encryption scheme is said to be additively homomorphic if for any two plaintexts $a, b \in \mathbb{M}$ (the message space), and their respective encryptions $Enc(a), Enc(b) \in \mathbb{C}$ (the ciphertext space) under the public key, there exists a group operation $\odot$ in $\mathbb{C}$ such that:
\[
Dec(Enc(a) \odot Enc(b)) = a + b
\]
where $Dec$ is the decryption function, and `+' is the addition operation in the plaintext space $\mathbb{M}$.
\end{definition}

\subsection{Homomorphic Computation of Inner Product}

The homomorphic computation of the inner product can be described as follows. Let $Enc$ denote the encryption function with $\odot$ as the operation in the ciphertext space, which may not be standard multiplication but rather a specific group operation defined for the encryption scheme. 
For some schemes, such as Pailliar~\cite{ppail_eurocrypt99}, the $\odot$ operation is indeed arithmetic multiplication.

The encrypted similarity, $Sim(\cdot, \cdot)$ of vectors \textbf{x} and $Enc(\textbf{y})$ = $\left( Enc(y_i) \right)$ can be computed as:
\[
Sim(\textbf{x}, Enc(\textbf{y})) = Enc\left(\sum_{i=1}^{d} x_i y_i\right) = \bigotimes_{i=1}^{d} Enc(y_i)^{x_i},
\]
where $\bigotimes$ denotes the multiplication of the factors and the exponentiation is realized by the multiplication of the base ciphertexts.
That is, 
\[
Enc(y_i)^{x_i} = \bigotimes_{j=1}^{x_i} Enc(y_i).
\]
Combining the above two equations, we have 
\begin{equation}
    Sim(\textbf{x}, Enc(\textbf{y})) = \bigotimes_{i=1}^{d} \bigotimes_{j=1}^{x_i} Enc(y_i),
\end{equation}
which allows us to compute the encrypted similarity between a pair of plaintext-ciphertext vectors.

\subsection{Correctness of the Homomorphic Inner Product Computation}

We prove that the encrypted similarity computation preserves the correctness of the inner product under decryption. Given a plaintext vector $\mathbf{x} = (x_1, x_2, \dots, x_d)$ and an encrypted vector $Enc(\mathbf{y}) = (Enc(y_1), Enc(y_2), \dots, Enc(y_d))$, the encrypted similarity is computed as:

\begin{equation}
    Sim(\mathbf{x}, Enc(\mathbf{y})) = \bigotimes_{i=1}^{d} \bigotimes_{j=1}^{x_i} Enc(y_i).
\end{equation}

Using the plaintext-ciphertext multiplication property of the encryption scheme, we have:

\begin{equation}
    Enc(y_i)^{x_i} = \bigotimes_{j=1}^{x_i} Enc(y_i) = Enc(x_i y_i).
\end{equation}

Applying this property to the inner product computation:

\begin{equation}
    Sim(\mathbf{x}, Enc(\mathbf{y})) = \bigotimes_{i=1}^{d} Enc(y_i)^{x_i} = \bigotimes_{i=1}^{d} Enc(x_i y_i).
\end{equation}

Since the encryption function $Enc$ is additively homomorphic, we use the homomorphic addition property:

\begin{equation}
    \bigotimes_{i=1}^{d} Enc(x_i y_i) = Enc\left(\sum_{i=1}^{d} x_i y_i\right).
\end{equation}

Applying the decryption function $Dec$, we obtain:

\begin{equation}
    Dec(Sim(\mathbf{x}, Enc(\mathbf{y}))) = Dec\left(Enc\left(\sum_{i=1}^{d} x_i y_i\right)\right).
\end{equation}

Since decryption correctly recovers the original plaintext:

\begin{equation}
    Dec\left(Enc\left(\sum_{i=1}^{d} x_i y_i\right)\right) = \sum_{i=1}^{d} x_i y_i.
\end{equation}

Thus, we have proven that the homomorphic inner product computation preserves correctness:

\begin{equation}
    Dec(Sim(\mathbf{x}, Enc(\mathbf{y}))) = \sum_{i=1}^{d} x_i y_i.
\end{equation}

This demonstrates that the proposed method allows secure computation of the encrypted similarity while ensuring correctness upon decryption.

\subsection{Security Analysis}

We analyze the security of the proposed homomorphic inner product computation under the standard semantic security model. Specifically, we show that if the underlying encryption scheme is semantically secure under chosen plaintext attacks (IND-CPA), then the encrypted similarity computation does not reveal any information about the encrypted vector $\mathbf{y}$.

\subsubsection{Threat Model}
We assume an adversary who observes all computations performed during the similarity computation process, including:
\begin{itemize}
    \item The plaintext vector $\mathbf{x}$, which is publicly known.
    \item The set of ciphertexts $\{ Enc(y_1), Enc(y_2), \dots, Enc(y_d) \}$.
    \item The final encrypted similarity result $Enc(s)$, where $s = \sum_{i=1}^{d} x_i y_i$.
\end{itemize}
The adversary's goal is to extract information about the encrypted values $\mathbf{y} = (y_1, y_2, \dots, y_d)$ beyond what is trivially deducible from $\mathbf{x}$ and $Enc(s)$.

\subsubsection{Reduction to the Semantic Security of the Encryption Scheme}

We prove security by reduction: if an adversary $\mathcal{A}$ can break the privacy of $\mathbf{y}$ in our similarity computation scheme, then we can construct a simulator $\mathcal{S}$ that uses $\mathcal{A}$ as a subroutine to break the semantic security (IND-CPA) of the underlying encryption scheme.

\begin{enumerate}
    \item Assume $\mathcal{A}$ can infer partial or full information about the encrypted values $y_i$ from observing the computation of $Enc(s)$.
    \item We construct $\mathcal{S}$ that interacts with an encryption oracle of an IND-CPA secure encryption scheme. The simulator $\mathcal{S}$:
    \begin{itemize}
        \item Chooses a random plaintext vector $\mathbf{x}$.
        \item Queries the encryption oracle to obtain $Enc(y_1), \dots, Enc(y_d)$ for randomly chosen $y_i$ values.
        \item Computes $Enc(s) = \bigotimes_{i=1}^{d} Enc(y_i)^{x_i}$.
    \end{itemize}
    \item If $\mathcal{A}$ can successfully extract information about any $y_i$, then $\mathcal{S}$ can use this to distinguish between encryptions of different messages, violating the IND-CPA security of the underlying encryption scheme.
\end{enumerate}

Since the encryption scheme is assumed to be IND-CPA secure, no polynomial-time adversary can extract any meaningful information about $y_i$ from its ciphertext $Enc(y_i)$. This ensures that our similarity computation process does not reveal additional information beyond what is trivially deducible from $\mathbf{x}$ and $Enc(s)$.

\subsection{Computational Complexity Analysis}

We analyze the computational complexity of the privacy-preserving inner product computation.

The total complexity consists of three main components: homomorphic multiplications (plaintext-ciphertext multiplications, PCM), homomorphic additions, and a final decryption step. Let $N$ denote the security parameter of the encryption scheme, typically corresponding to the bit-length of the modulus. From the inner product computation formula:

\begin{equation}
    Sim(\mathbf{x}, Enc(\mathbf{y})) = \bigotimes_{i=1}^{d} \bigotimes_{j=1}^{x_i} Enc(y_i),
\end{equation}

we derive the following computational costs:

\begin{itemize}
    \item \textbf{Plaintext-Ciphertext Multiplication (PCM) Cost:} Each encrypted value $Enc(y_i)$ is exponentiated by $x_i$, requiring $\mathcal{O}(\log x_i)$ modular multiplications using square-and-multiply exponentiation. For all $d$ dimensions, the total cost is $\mathcal{O} ( d \cdot T_{PCM} )$.

    \item \textbf{Homomorphic Addition Cost:} Since we perform $d-1$ additions in the encrypted domain, the total cost is $\mathcal{O} ( d \cdot T_{Add} )$.

    \item \textbf{Decryption Cost:} Only one final decryption is performed, contributing $\mathcal{O} ( T_{Dec} )$.
\end{itemize}

Thus, the overall computational complexity is:

\begin{equation}
    \mathcal{O} \left( d \cdot (T_{PCM} + T_{Add}) + T_{Dec} \right).
\end{equation}

For encryption schemes supporting additive homomorphism, we estimate:
\begin{align*}
    T_{PCM} & = \mathcal{O}(\log x_i \cdot \log^2 N), \\
    T_{Add} & = \mathcal{O}(1), \\
    T_{Dec} & = \mathcal{O}(\log^2 N).
\end{align*}

Thus, the total complexity simplifies to:

\begin{equation}
    \mathcal{O} \left( d \cdot \log^2 N \right).
\end{equation}

Compared to fully homomorphic encryption (FHE), which typically requires $\mathcal{O}(\text{poly}(\log N))$ per operation due to bootstrapping overhead, this approach is significantly more efficient, making it suitable for real-time privacy-preserving similarity computation.

\subsection{Error Growth in Homomorphic Inner Product Computation}

In homomorphic encryption schemes, noise accumulates with each operation, impacting the correctness of the decryption process. For additively homomorphic encryption (AHE), the primary operations involved in computing an encrypted inner product are homomorphic additions and plaintext-ciphertext multiplications. We analyze the error growth in this context and discuss its implications for practical encrypted similarity search.

\paragraph{Noise Growth in Addition.} 
Let \( \mathsf{Enc}(y_i) \) denote the encryption of a plaintext value \( y_i \), and assume the encryption scheme introduces an initial noise term \( e_i \) such that:
\[
\mathsf{Enc}(y_i) = y_i + e_i \mod q.
\]
For an encrypted inner product:
\[
\mathsf{Enc}(s) = \sum_{i=1}^{d} x_i \cdot \mathsf{Enc}(y_i),
\]
the resulting noise term is:
\[
e_s = \sum_{i=1}^{d} x_i e_i.
\]
If \( e_i \) follows a uniform distribution \( U(-B, B) \), then the variance of \( e_s \) is:
\[
\text{Var}(e_s) = \sum_{i=1}^{d} x_i^2 \cdot \text{Var}(e_i) = d B^2 \mathbb{E}[x_i^2].
\]
Thus, the noise grows linearly with dimension \( d \), which is manageable for moderate dimensions but may require parameter tuning for high-dimensional vectors.

\paragraph{Asymptotic Noise Growth Bound.} 
For a worst-case analysis, assuming \( x_i \) is bounded by \( X \), we obtain:
\[
|e_s| \leq d X B.
\]
If \( X \) and \( B \) are constants, the noise growth remains \( O(d) \), ensuring scalability for practical applications. However, for large plaintext values \( X \), the noise can accumulate rapidly, requiring modulus switching or rescaling to maintain decryption accuracy.

\paragraph{Comparison to FHE.} 
Unlike fully homomorphic encryption (FHE), where noise grows exponentially due to ciphertext-ciphertext multiplications, AHE only experiences linear noise growth. This eliminates the need for bootstrapping, significantly reducing computation cost. Instead, AHE can use modulus switching or rescaling to maintain correctness over multiple operations while keeping the noise under control.

\section{Conclusion}

This work investigates the feasibility of performing privacy-preserving similarity search using additively homomorphic encryption (AHE). While fully homomorphic encryption (FHE) enables arbitrary computations over encrypted data, its high computational overhead limits its practicality in large-scale applications. In contrast, AHE provides a lightweight alternative, supporting addition and scalar multiplication while significantly reducing computational costs.

We show that inner product similarity, a fundamental metric in vector retrieval, can be efficiently computed under AHE while maintaining privacy and correctness. Our approach is particularly well-suited for privacy-sensitive applications where only one component (either the database or the query vector) needs encryption, such as secure federated learning and confidential recommendation systems.

Future work will explore optimizations for specific encryption schemes, potential extensions to other similarity metrics, and trade-offs between efficiency and security in real-world deployments.

\bibliographystyle{splncs04}
\bibliography{mybibliography}

\end{document}